\documentclass[twocolumn,nopacs,floatfix,amsmath,nofootinbib,amssymb,preprintnumbers,floatfix]{revtex4}
\usepackage{longtable,lscape,booktabs}
\usepackage{txfonts}
\usepackage{overpic}
\usepackage{amssymb}
\usepackage{indentfirst}
\usepackage{feynmf}
\usepackage{slashed}
\usepackage{cases}
\usepackage{color}
\usepackage{multirow}
\usepackage{ulem}
\usepackage{subfigure}
\usepackage{graphicx,color,dcolumn,bm}
\usepackage[colorlinks,
            citecolor=blue,
            anchorcolor=red,
            menucolor=red,
            linkcolor=red,
            filecolor=red,
            runcolor=red,
            urlcolor=blue,
            frenchlinks=red]{hyperref}

\begin{document}
\title{Color gauge invariant theory of diquark interactions}
\author{Jun-Feng Wang$^{1,2}$}
\email[]{wjunfeng2023@lzu.edu.cn}	
\author{De-Shun Zhang$^{1,2}$}
\email[]{220220940071@lzu.edu.cn}
\author{Zhi-Feng Sun$^{1,2,3,4}$}
\email[Corresponding author: ]{sunzf@lzu.edu.cn}
	
\affiliation {\it$^1$School of Physical Science and Technology, Lanzhou University, Lanzhou 730000, China\\
$^2$Research Center for Hadron and CSR Physics, Lanzhou University and Institute of Modern Physics of CAS, Lanzhou 730000, China\\
$^3$Lanzhou Center for Theoretical Physics, Key Laboratory of Theoretical Physics of Gansu Province, Key Laboratory of Quantum Theory and Applications of MoE, Gansu Provincial Research Center for Basic Disciplines of Quantum Physics, Lanzhou University, Lanzhou 730000, China\\
$^4$Frontiers Science Center for Rare Isotopes, Lanzhou University, Lanzhou, Gansu 730000, China}
	
\date{\today}
	
\begin{abstract}
In the present work, we construct a color gauge invariant theory of diquark interactions. With the transformation rule of the diquark fields and the definition of the covariant derivatives under color SU(3) symmetry, we construct the gauge invariant Lagrangians describing the vertices of $S-S-G$, $A-A-G$, $S_Q-S_Q-G$, $A_Q-A_Q-G$ and $S_Q-A_Q-G$, where $S_{(Q)}$ is the light (heavy) scalar diquark field, $A_{(Q)}$ is the light (heavy) axial vector diquark field, and $G$ is the gluon field. And then we derive the one-gluon-exchange effective potentials for diquark-antidiquark interactions using the obtained Lagrangians. By comparing these potentials with those in Godfrey-Isgur quark model, the coupling constants of the Lagrangians are determined. We find that the potentials are mainly made of Coulomb, contact and tensor terms. The potential for the process $S\bar{A}\to A\bar{S}$ is 0, since $\mathcal{L}_{SAG}=0$. And the tensor terms proportional to $g^2$ are negligible.
\end{abstract}
	
\maketitle
	
\section{Introduction}\label{I}
Before 2003, there exist only a few candidates of exotic states, e.g. the light scalar mesons. At that time, the studies were far from satisfactory due to the constraint of reliable experimental and theoretical techniques, and no candidates for states beyond ordinary hadrons containing heavy quark(s) had been found. So the topics of exotic states had been low-profile for a long time. In 2003, the situation changed when several important candidates of exotic states were discovered, which includes the $X(3872)/\chi_{c1}(3872)$ observed by Belle, the $D_{s0}^{*}(2317)$ observed by BaBar, and the $D_{s1}(2460)$ observed by CLEO. Since then more and more exotic candidates were continually observed in the high energy experiments, such as BaBar, Belle, BESIII, LHCb, and so on. In order to explain these states, many models have been proposed, e.g. molecular states, diquark-antidiquark states, kinematical effects and so on \cite{Chen:2016qju,Liu:2019zoy,Drenska:2010kg,Olsen:2014qna,Hosaka:2016pey,Briere:2016puj,Richard:2016eis,Lebed:2016hpi,Esposito:2016noz,Guo:2017jvc,Ali:2017jda,Karliner:2017qhf,Albuquerque:2018jkn,Brambilla:2019esw,Guo:2019twa,Meng:2022ozq,Bicudo:2022cqi,Johnson:2024omq,Liu:2024uxn,Chen:2022asf,Dong:2021bvy,Dong:2017gaw,Yang:2020atz,Mezzadri:2022loq}.

In our previous works \cite{Cao:2022rjp,He:2024aej,Zhang:2024zbo,Ma:2024zqf,Zhang:2025}, we proposed a scheme in which the mixture effect between molecular state and diquark-antidiquark state within the effective field theory is investigated. There, we construct the Lagrangians describing heavy diquark-light diquark-heavy meson, heavy diquark-heavy diquark-light meson, light diquark-light diquark-light meson vertices under the hidden local symmetry. The theory indicates that the $Z_{cs}(4000)^{+}$, $Z_{cs}(4220)^{+}$, $Z_{c}(3900)$, $X(4500)$ and $T_{c\Bar{s}0}^{*}(2900)$ can be explained as the mixture of hadronic molecule and diquark-antidiquark state, and $Z_{b}(10610)$ and $Z_{b}(10650)$ as molecular states slightly mixing with diquark-antidiquark states.

However, the interactions of diquarks with gluon have not been considered in these works. 

It is known that the establishment of U(1) gauge symmetry successfully described the electromagnetic interaction. In 1954, Chen Ning Yang and Robert L. Mills extended the idea of gauge invariance to non-abelian group, i.e., SU(2) group, in their epoch-making paper \cite{Yang:1954ek}, which led to the birth of the electroweak theory and quantum chromodynamics (QCD). In the present work, we further extend the SU(3) color gauge symmetry to the diquark sector, and construct a color gauge invariant theory of diquark interactions in which the gluon is introduced as the gauge boson.

The layout of the paper is as follows. After the introduction, we systematically construct color gauge invariant Lagrangians in Sec.~\ref{II}. The one-gluon-exchange potentials of diquark-antidiquark interactions obtained from the constructed Lagrangians are given in Sec.~\ref{III}. In Sec.~\ref{IV} comparing with the obtained potentials with those deduced from the Godfrey-Isgur quark model, the couplings of diquark-diquark-gluon vertices are determined. A short summary is given finally.

\section{The construction of color gauge invariant Lagrangians}\label{II}
In this work, we study the interactions of diquarks and gloun. So we need to construct the Lagrangians which is SU(3)$_{c}$ gauge invariant. In order to simplify the research, we take the assumption that the diquark is a point-like particle. 

Since a diquark is made up of two quarks, its color structure can be classified into antitriplet and sextet, according to group theory, i.e., $3\otimes3=\Bar{3}\oplus6$. The interaction between the two quarks of a color antitriplet is attractive, while that of a color sextet is repulsive. So we only consider color antitriplet diquarks in the present work. The diquark fields in the flavour space can be introduced in the matrix form as follows
\begin{eqnarray}
    S_{F}&=&\left(
    \begin{array}{ccc}
        0      &    S_{ud}    &S_{us}  \\
    -S_{ud}    &      0       &S_{ds}  \\
    -S_{us}    &   -S_{ds}    &0
    \end{array}\right),\label{eq1} \\
    A_{F}&=&\left(
    \begin{array}{ccc}
            A_{uu}             &  \frac{1}{\sqrt{2}}A_{ud}  & \frac{1}{\sqrt{2}}A_{us}  \\
    \frac{1}{\sqrt{2}}A_{ud}   &          A_{dd}            &  \frac{1}{\sqrt{2}}A_{ds}  \\
    \frac{1}{\sqrt{2}}A_{us}   &  \frac{1}{\sqrt{2}}A_{ds}  &          A_{ss}
    \end{array}\right),\label{eq2} \\
    S_{QF}&=&\left(S_{Qu},S_{Qd},S_{Qs}\right),\label{eq3} \\
    A_{QF}&=&\left(A_{Qu},A_{Qd},A_{Qs}\right).\label{eq4}
\end{eqnarray}
Here, $S_{F}$ and $S_{QF}$ denote scalar diquarks, $A_{F}$ and $A_{QF}$ denote axial vector diquarks. The subscript $Q$ corresponds to heavy quark $c$ or $b$, and $F$ means the matrix is in the flavour space. In the light diquark sector, the two quarks are identical. For a color antitriplet, the color wave function is antisymmetric. In this work, only ground state diquarks are taken into account. So the spatial wave function is symmetric. The spin wave functions of scalar and axial vector diquarks are antisymmetric and symmetric, respectively. As a result, the flavor wave functions for scalar and axial vector diquarks are antisymmetric and symmetric, respectively. In the color space, the diquark fields can be written as
\begin{eqnarray}
    X_{C}=\left(
    \begin{array}{ccc}
         0      &    X_{rg}    &  X_{rb}  \\
     -X_{rg}    &      0       &  X_{gb}  \\
     -X_{rb}    &   -X_{gb}    &   0
    \end{array}\right)\label{eq5}
\end{eqnarray}
with $X=S,A,S_{Q}$ and $A_{Q}$, the subscript $C$ denoting the matrix is in the color space.

The free Lagrangian terms of light diquarks and heavy-light diquarks are listed as follows:
\begin{eqnarray}
\mathcal{L}_{S}&=&\frac{1}{4}\langle \partial_{\mu}S\partial^{\mu}S^{\dagger}\rangle _{CF}-\frac{1}{4}m_{S}^{2}\langle SS^{\dagger}\rangle _{CF},\label{eq6} \\
\mathcal{L}_{A}&=&-\frac{1}{4}\langle F_{\mu\nu}F^{\dagger\mu\nu}\rangle _{CF}+\frac{1}{2}m_{A}^{2}\langle A_{\mu}A^{\dagger\mu}\rangle _{CF},\label{eq7} \\
\mathcal{L}_{S_{Q}}&=&\frac{1}{2}\langle \partial_{\mu}S_{Q}\partial^{\mu}S^{\dagger}_{Q}\rangle _{C}-\frac{1}{2}m_{S_{Q}}^{2}\langle S_{Q}S^{\dagger}_{Q}\rangle _{C},\label{eq8} \\
\mathcal{L}_{A_{Q}}&=&-\frac{1}{4}\langle F_{Q\mu\nu}F^{\dagger\mu\nu}_{Q}\rangle _{C}+\frac{1}{2}m_{A_{Q}}^{2}\langle A_{Q\mu}A^{\dagger\mu}_{Q}\rangle _{C}.\label{eq9} 
\end{eqnarray}
Here, $F_{\mu\nu}=\partial_{\mu}A_{\nu}-\partial_{\nu}A_{\mu}$ and $F_{Q\mu\nu}=\partial_{\mu}A_{Q\nu}-\partial_{\nu}A_{Q\mu}$ are the strength tensor of the axial vector diquark fields, $\langle \cdots\rangle _{CF}$ denotes the trace in both color and flavor space, and $\langle \cdots\rangle _{C}$ is the trace only in color space. Under SU(3)$_{c}$ gauge transformation, the diquark fields transforms by
\begin{eqnarray}
S_{C} &\rightarrow& US_{C}U^{T},\label{eq10} \\
A_{C\mu} &\rightarrow& UA_{C\mu}U^{T},\label{eq11} \\
S_{QC} &\rightarrow& US_{QC}U^{T},\label{eq12} \\
A_{QC\mu} &\rightarrow& UA_{QC\mu}U^{T},\label{eq13}
\end{eqnarray}
where $U$ is the element of the SU(3) group in color space. If the transformation is global, the Lagrangians in Eqs. \eqref{eq6}-\eqref{eq9} are invariant. However, if the transformation is local, i.e., $U$ depends on spacetime coordinates, the invariance of these Lagrangians is violated. In order to maintain the symmetry of Lagrangians under the local gauge transformation, in analogy with the QCD case, we need to introduce a gauge field, i.e., gluon field, whose transformation rule is given by
\begin{eqnarray}
G_{\mu}\equiv G^{a}_{\mu}t^{a} &\rightarrow& U(G^{a}_{\mu}t^{a}+\frac{i}{g}\partial_{\mu})U^{\dag}.\label{eq14}
\end{eqnarray}
Here, $t^{a}=\frac{\lambda^{a}}{2}$ with $\lambda^{a} (a=1,\cdots 8)$ the Gell-Mann matrix. The strength tensor of the gluon field $G_{\mu\nu}^{a}=\partial_{\mu}G_{\nu}^{a}-\partial_{\nu}G_{\mu}^{a}+gf^{abc}G_{\mu}^{b}G_{\nu}^{c}$ transforms by $G_{\mu\nu}^{a}\rightarrow UG_{\mu\nu}^{a}U^{\dagger}$. The covariant derivatives for diquark fields are defined as follows
\begin{eqnarray}
D_\mu S&=&\partial_{\mu}S-iV_{\mu}S-iSV_{\mu}^{T}-igG_{\mu}S-igSG_{\mu}^{T},\label{eq15} \\ 
D_{\mu}A_{\nu}&=&\partial_{\mu}A_{\nu}-iV_{\mu}A_{\nu}-iA_{\nu}V_{\mu}^{T}-igG_{\mu}A_{\nu}-igA_{\nu}G_{\mu}^{T},\label{eq16} \\ 
D_\mu S_{Q}&=&\partial_{\mu}S_{Q}-iS_{Q}\alpha_{\parallel \mu}^{T}-igG_{\mu}S_{Q}-igS_{Q}G_{\mu}^{T},\label{eq17}\\
D_{\mu}A_{Q\nu}&=&\partial_{\mu}A_{Q\nu}-iA_{Q\nu}\alpha_{\parallel \mu}^{T}-igG_{\mu}A_{Q\nu}-igA_{Q\nu}G_{\mu}^{T}.\label{eq18}
\end{eqnarray}
which transforms in the color space by
\begin{eqnarray}
D_{\mu}S &\rightarrow& UD_{\mu}SU^{T},\label{eq19} \\
D_{\mu}A_{\nu} &\rightarrow& UD_{\mu}A_{\nu}U^{T},\label{eq20} \\
D_{\mu}S_{Q} &\rightarrow& UD_{\mu}S_{Q}U^{T},\label{eq21} \\
D_{\mu}A_{Q\nu} &\rightarrow& UD_{\mu}A_{Q\nu}U^{T}.\label{eq22}
\end{eqnarray}
In Eqs. \eqref{eq15}-\eqref{eq18}, 
\begin{eqnarray}
V_\mu &=&\frac{g_{V}}{\sqrt{2}}\left(
\begin{array}{ccc}
\frac{1}{\sqrt{2}}(\rho^{0}+\omega)&\rho^+&K^{*+}\\
\rho^-&-\frac{1}{\sqrt{2}}(\rho^{0}-\omega)&K^{*0}\\
K^{*-}&\bar{K}^{*0}&\phi
\end{array}
\right)_\mu,\\
\alpha_\parallel^\mu&=&(\partial_{\mu}\xi_{R}\xi_{R}^{\dag}+\partial_{\mu}\xi_{L}\xi_{L}^{\dag})/2i,  
\end{eqnarray}
with
\begin{eqnarray}
\xi_{L,R}&=&e^{i\sigma/f_\sigma}e^{\mp i\Phi/(2f_\pi)},\\
\Phi&=&\left(
\begin{array}{ccc}
\frac{\sqrt{3}\pi^0+\eta+\sqrt{2}\eta'}{\sqrt{3}}&\sqrt{2}\pi^+&\sqrt{2}K^+\\
\sqrt{2}\pi^-&\frac{-\sqrt{3}\pi^0+\eta+\sqrt{2}\eta'}{\sqrt{3}}&\sqrt{2}K^0\\
\sqrt{2}K^-&\sqrt{2}\bar{K}^0&\frac{-2\eta+\sqrt{2}\eta'}{\sqrt{3}}
\end{array}
\right),
\end{eqnarray}
which is introduced under the hidden local symmetry in the flavor space. The covariant field strength tensor for the axial vector diquark fields read
\begin{eqnarray}
    \widetilde{F}_{\mu\nu}&\equiv&D_{\mu}A_{\nu}-D_{\nu}A_{\mu}, \label{eq27} \\
    \widetilde{F}_{Q\mu\nu}&\equiv&D_{\mu}A_{Q\nu}-D_{\nu}A_{Q\mu}, \label{eq28}
\end{eqnarray}
with the following transformation rule
\begin{eqnarray}
    \widetilde{F}_{\mu\nu}&\rightarrow&U\widetilde{F}_{\mu\nu}U^{T}, \label{eq29}\\
    \widetilde{F}_{Q\mu\nu}&\rightarrow&U\widetilde{F}_{Q\mu\nu}U^{T}, \label{eq30}
\end{eqnarray}
with the above presentions, we construct the color gauge invariant Lagrangians which are shown as following
\begin{eqnarray}
\mathcal{L}'_{S}&=&\frac{1}{4}\langle D_{\mu}SD^{\mu}S^{\dagger}\rangle_{CF} -\frac{1}{4}m_{S}^{2}\langle SS^{\dagger}\rangle_{CF} ,\label{eq31} \\
\mathcal{L}'_{A}&=&-\frac{1}{4}\langle \widetilde{F}_{\mu\nu}\widetilde{F}^{\dagger\mu\nu}\rangle_{CF} +\frac{1}{2}m_{A}^{2}\langle A_{\mu}A^{\dagger\mu}\rangle_{CF} ,\label{eq32} \\
\mathcal{L}'_{S_{Q}}&=&\frac{1}{2}\langle D_{\mu}S_{Q}D^{\mu}S^{\dagger}_{Q}\rangle_{C} -\frac{1}{2}m_{S_{Q}}^{2}\langle S_{Q}S^{\dagger}_{Q}\rangle_{C} ,\label{eq33} \\
\mathcal{L}'_{A_{Q}}&=&-\frac{1}{4}\langle \widetilde{F}_{Q\mu\nu}\widetilde{F}^{\dagger\mu\nu}_{Q}\rangle_{C} +\frac{1}{2}m_{A_{Q}}^{2}\langle A_{Q\mu}A^{\dagger\mu}_{Q}\rangle_{C} .\label{eq34} 
\end{eqnarray}

The existence of $S_{Q}-A_{Q}-G$, $A-A-G$ and $A_{Q}-A_{Q}-G$ vertices is also possible. The expressions of them read
\begin{eqnarray}
\mathcal{L}_{AAG}&=&id_2\langle A^{\mu}G_{\mu\nu}^{T}A^{\dagger\nu}\rangle_{CF},\label{eq35}\\
\mathcal{L}_{S_{Q}A_{Q}G}&=&d_{1Q}\epsilon^{\mu\nu\alpha\beta}\langle D_{\mu}S_{Q}G_{\nu\alpha}^{T}A_{Q\beta}^{\dagger}+A_{Q\beta}G_{\nu\alpha}^{T}D_{\mu}S_{Q}^{\dagger}\rangle_{C},\label{eq36}\\
\mathcal{L}_{A_{Q}A_{Q}G}&=&id_{2Q}\langle A_Q^{\mu}G_{\mu\nu}^{T}A_Q^{\dagger\nu}\rangle_{C}.\label{eq37}
\end{eqnarray}
Note here that there is no Lagrangian describing the vertex of $S-A-G$, i.e., $\mathcal{L}_{SAG}=0$, because the diquark field $S$ is antisymmetric and $A$ is symmetric in the flavour space. This is discussed in details in Appendix B. However, for $\mathcal{L}_{S_QA_QG}$, $S_Q$ and $A_Q$ do not carry such symmetry. So $\mathcal{L}_{S_QA_QG}$ does not vanish. Besides, $\mathcal{L}_{S_{Q}A_{Q}G}$ do not involve the terms proportional to
\begin{eqnarray}
    \epsilon^{\mu\nu\alpha\beta}\langle S_{Q}G_{\mu\nu}^{T}F_{Q\alpha\beta}^{\dagger}\rangle_{C} +h.c. , \label{eq38}
\end{eqnarray}
since it is equivalent to $\mathcal{L}_{S_{Q}A_{Q}G}$. The proof is shown in Appendix A. 

The Lagrangians corresponding to light diquark-light diquark-light meson, heavy diquark-heavy meson-light diquark and heavy diquark-heavy diquark-light meson vertices have already been constructed in Refs. \cite{Cao:2022rjp,He:2024aej,Zhang:2024zbo,Ma:2024zqf,Zhang:2025}. If we want to get color gauge invariant Lagrangians, we need to replace the covariant derivatives defined in Refs. \cite{Cao:2022rjp,He:2024aej,Zhang:2024zbo,Ma:2024zqf,Zhang:2025} with the ones in Eqs. \eqref{eq15}-\eqref{eq18}, and add a few terms not containing covariant derivative but $G_{\mu\nu}$:
\begin{eqnarray}
\mathcal{L}_1&=&i\Tilde{k}_{1}\langle S\hat{\alpha}_{\|}^{\mu T}D_{\mu}S^{\dag} \rangle_{CF}-i\Tilde{k}_{1}\langle D_{\mu}S\hat{\alpha}_{\|}^{\mu T}S^{\dag}\rangle_{CF} \nonumber\\
&&+i\Tilde{k}_{2}\langle S\hat{\alpha}_{\bot\mu}^{T}A^{\mu\dag}\rangle_{CF} -i\Tilde{k}_{2}\langle A^{\mu}\hat{\alpha}_{\bot\mu}^{T}S^{\dag}\rangle_{CF} \nonumber\\
&&+i\Tilde{k}_{3}\langle \Tilde{F}_{\mu\nu}\hat{\alpha}_{\|}^{\mu T}A^{\nu\dag}\rangle_{CF} -i\Tilde{k}_{3}\langle A_{\nu}\hat{\alpha}_{\|\mu}^{T}\Tilde{F}^{\mu\nu\dag}\rangle_{CF} \nonumber\\
&&+\Tilde{k}_{4}\epsilon^{\mu\nu\alpha\beta}\langle \Tilde{F}_{\mu\nu}\hat{\alpha}_{\bot\alpha}^{T}A_{\beta}^{\dag}\rangle_{CF} +\Tilde{k}_{4}\epsilon^{\mu\nu\alpha\beta}\langle A_{\beta}\hat{\alpha}_{\bot\alpha}^{T}\Tilde{F}_{\mu\nu}^{\dag}\rangle_{CF},\label{eq39}\nonumber\\
\end{eqnarray}
\begin{eqnarray}
\mathcal{L}_2&=&\Tilde{e}_{1}\langle iPD_{\mu}SA_{Q}^{\mu\dag}-iA_{Q}^{\mu}D_{\mu}S^{\dag}P^{\dag}\rangle_{C}\nonumber\\
&&+\Tilde{e}_{2}\langle iPA_{\mu}D^{\mu}S_{Q}^{\dag}-iD^{\mu}S_{Q}A_{\mu}^{\dag}P^{\dag}\rangle_{C}\nonumber\\
&&+\Tilde{e}_{3}\langle\epsilon^{\mu\nu\alpha\beta}P\Tilde{F}_{\mu\nu}\Tilde{F}_{Q\alpha\beta}^{\dag}+\epsilon^{\mu\nu\alpha\beta}\Tilde{F}_{Q\alpha\beta}\Tilde{F}_{\mu\nu}^{\dag}P^{\dag}\rangle_{C}\nonumber\\
&&+\Tilde{e}_{4}\langle iP_{\mu}^{*}D^{\mu}SS_{Q}^{\dag}-iS_{Q}D^{\mu}S^{\dag}P_{\mu}^{*\dag}\rangle_{C}\nonumber\\
&&+\Tilde{e}_{5}\langle\epsilon^{\mu\nu\alpha\beta}P_{\mu}^{*}D_{\nu}S\Tilde{F}_{Q\alpha\beta}^{\dag}+\epsilon^{\mu\nu\alpha\beta}\Tilde{F}_{Q\alpha\beta}D_{\nu}S^{\dag}P_{\mu}^{*\dag}\rangle_{C}\nonumber\\
&&+\Tilde{e}_{6}\langle\epsilon^{\mu\nu\alpha\beta}P_{\mu}^{*}\Tilde{F}_{\nu\alpha}D_{\beta}S_{Q}^{\dag}+\epsilon^{\mu\nu\alpha\beta}D_{\beta}S_{Q}\Tilde{F}_{\nu\alpha}^{\dag}P_{\mu}^{*\dag}\rangle_{C}\nonumber\\
&&+\Tilde{e}_{7}\langle iP_{\mu}^{*}\Tilde{F}^{\mu\nu}A_{Q\nu}^{\dag}-iA_{Q\nu}\Tilde{F}^{\mu\nu\dag}P_{\mu}^{*\dag}\rangle_{C}\nonumber\\
&&+\Tilde{e}_{8}\langle iP_{\mu}^{*}A_{\nu}\Tilde{F}_{Q}^{\mu\nu\dag}-i\Tilde{F}_{Q}^{\mu\nu}A_{\nu}^{\dag}P_{\mu}^{*\dag}\rangle_{C}\nonumber\\
&&+\Tilde{e}_{9}\langle iP_{\mu\nu}^{*}A^{\mu}A_{Q}^{\nu\dag}-iA_{Q}^{\nu}A^{\mu\dag}P_{\mu\nu}^{*\dag}\rangle_{C}\nonumber\\
&&+\Tilde{e}_{10}\langle\epsilon^{\mu\nu\alpha\beta}PA_{\mu}G_{\nu\alpha}^TA_{Q\beta}^{\dag}+\epsilon^{\mu\nu\alpha\beta}A_{Q\beta}G_{\nu\alpha}^TA_{\mu}^{\dag}P^{\dag}\rangle_{C}\nonumber\\
&&+\Tilde{e}_{11}\langle\epsilon^{\mu\nu\alpha\beta}P_{\mu}^{*}SG_{\nu\alpha}^TA_{Q\beta}^{\dag}+\epsilon^{\mu\nu\alpha\beta}A_{Q\beta}G_{\nu\alpha}^TS^{\dag}P_{\mu}^{*\dag}\rangle_{C}\nonumber\\
&&+\Tilde{e}_{12}\langle\epsilon^{\mu\nu\alpha\beta}P_{\mu}^{*}A_{\nu}G_{\alpha\beta}^TS_{Q}^{\dag}+\epsilon^{\mu\nu\alpha\beta}S_{Q}G_{\alpha\beta}^TA_{\nu}^{\dag}P_{\mu}^{*\dag}\rangle_{C},\label{eq40}\\
\mathcal{L}_3&=&\Tilde{h}_{1}\langle iS_{Q}\hat{\alpha}_{\|}^{\mu T}D_{\mu}S_{Q}^{\dag}-iD_{\mu}S_{Q}\hat{\alpha}_{\|}^{\mu T}S_{Q}^{\dag}\rangle_{C}\nonumber\\
&&+\Tilde{h}_{2}\langle\epsilon^{\mu\nu\alpha\beta}\Tilde{F}_{Q\mu\nu}\hat{\alpha}_{\|\alpha}^{T}D_{\beta}S_{Q}^{\dag}+\epsilon^{\mu\nu\alpha\beta}D_{\beta}S_{Q}\hat{\alpha}_{\|\alpha}^{T}\Tilde{F}_{Q\mu\nu}^{\dag}\rangle_{C}\nonumber\\
&&+\Tilde{h}_{3}\langle iA_{Q\mu}\hat{\alpha}_{\bot}^{\mu T}S_{Q}^{\dag}-iS_{Q}\hat{\alpha}_{\bot}^{\mu T}A_{Q\mu}^{\dag}\rangle_{C}\nonumber\\
&&+\Tilde{h}_{4}\langle iA_{Q\mu}\hat{\alpha}_{\|\nu}^{T}\Tilde{F}_{Q}^{\mu\nu\dag}-i\Tilde{F}_{Q}^{\mu\nu}\hat{\alpha}_{\|\nu}^{T}A_{Q\mu}^{\dag}\rangle_{C}\nonumber\\
&&+\Tilde{h}_{5}\langle\epsilon^{\mu\nu\alpha\beta}A_{Q\mu}\hat{\alpha}_{\bot\nu}^{T}\Tilde{F}_{Q\alpha\beta}^{\dag}+\epsilon^{\mu\nu\alpha\beta}\Tilde{F}_{Q\alpha\beta}\hat{\alpha}_{\bot\nu}^{T}A_{Q\mu}^{\dag}\rangle_{C}\nonumber\\
&&+\Tilde{h}_{6}\langle\epsilon^{\mu\nu\alpha\beta}A_{Q\mu}G_{\nu\alpha}^T\hat{\alpha}_{\|\beta}^{T}S_{Q}^{\dag}+\epsilon^{\mu\nu\alpha\beta}S_{Q}\hat{\alpha}_{\|\beta}^{T}G_{\nu\alpha}^TA_{Q\mu}^{\dag}\rangle_{C}.\nonumber\\ \label{eq41}
\end{eqnarray}
In the present work, the definitions of the diquark fields are different from those in Refs. \cite{Cao:2022rjp,He:2024aej,Zhang:2024zbo,Ma:2024zqf,Zhang:2025}, resulting that the coupling constants in Eqs. \eqref{eq39}-\eqref{eq41} have the following relations to those in Refs. \cite{Cao:2022rjp,He:2024aej,Zhang:2024zbo,Ma:2024zqf,Zhang:2025}
\begin{eqnarray}
\Tilde{k}_i&=&-2k_i,\ (i=1, 2, 3, 4),\nonumber\\ 
\Tilde{e}_i&=&-2\sqrt{M_{m, i}M_{d, i}}e_i,\ (i=1,2,\cdots, 12),\nonumber\\
\Tilde{h}_i&=&-2\sqrt{M_{d,i}M^\prime_{d, i}}h_i,\ (i=1,2,\cdots, 6),
\end{eqnarray}
where $M_{m, i}$ and $M^{(\prime)}_{d, i}$ are the masses of heavy meson and heavy diquark in a certain term of the Lagrangians, respectively. 

\section{One-gluon-exchange potentials of diquark and antidiquark}\label{III}
\begin{figure}[h]
\subfigure[]{\label{FIG1a}
\scalebox{0.14}{\includegraphics{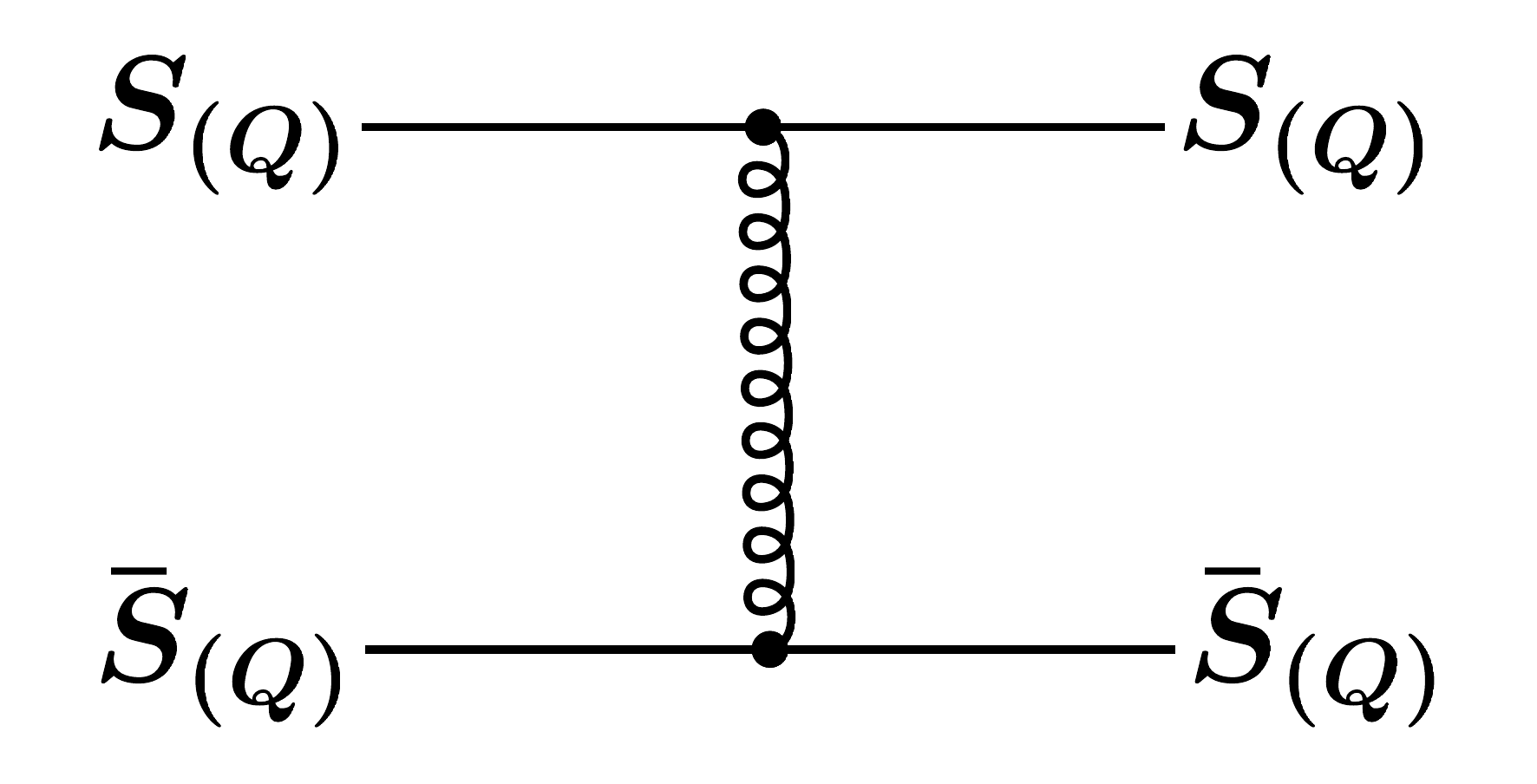}}} 
\subfigure[]{\label{FIG1b}
\scalebox{0.14}{\includegraphics{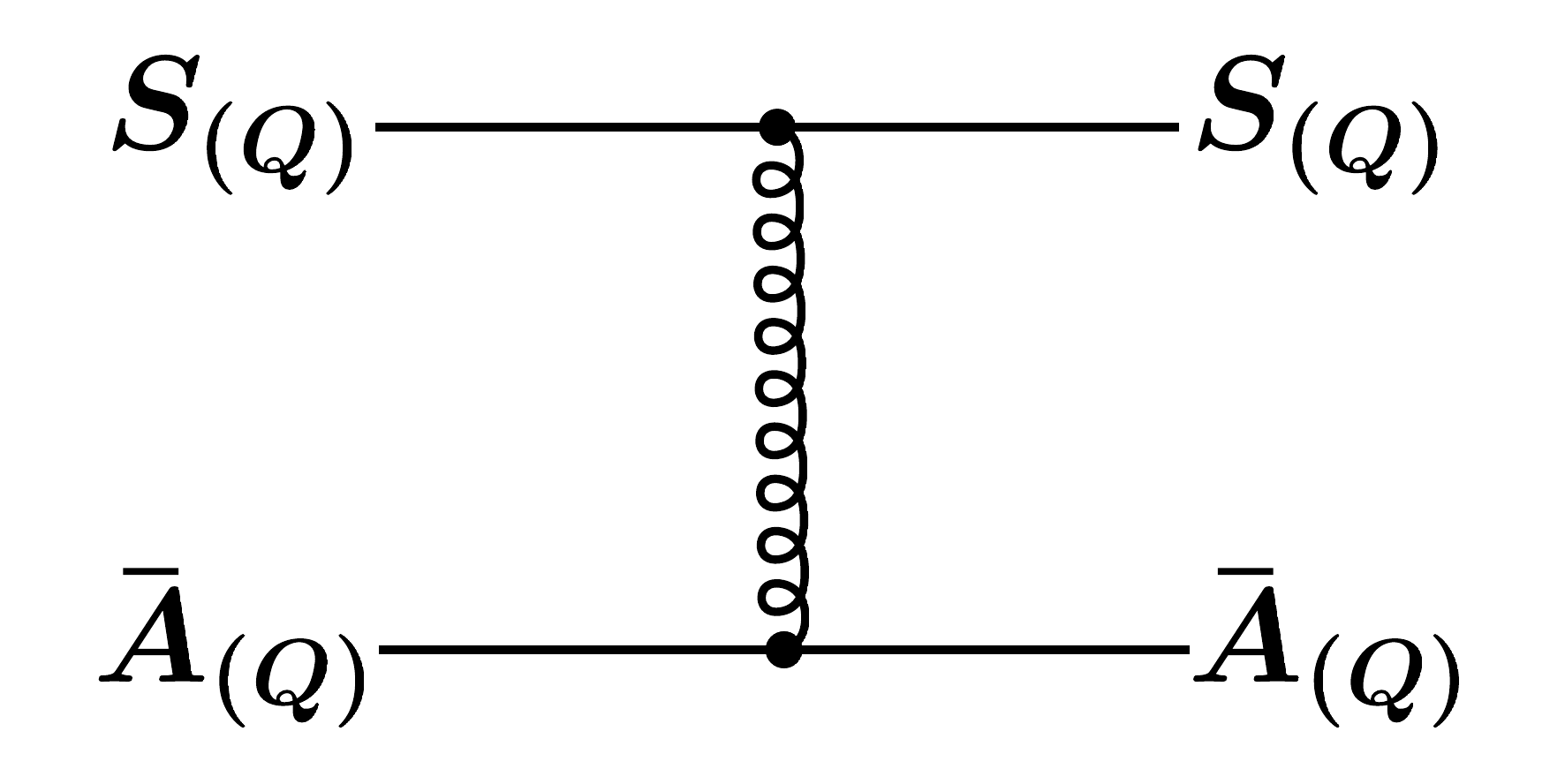}}} 
\\
\subfigure[]{\label{FIG1c}
\scalebox{0.14}{\includegraphics{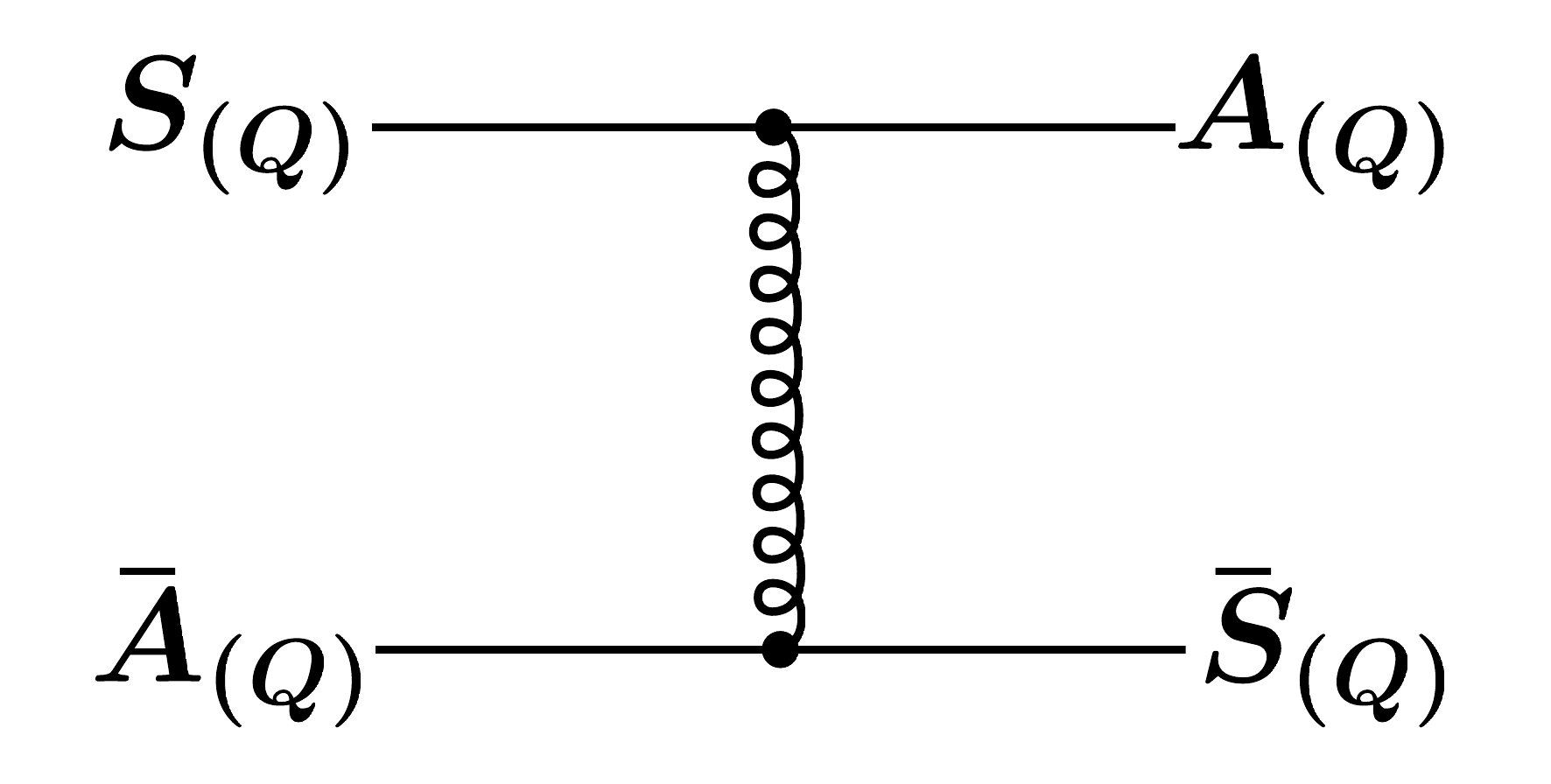}}}
\subfigure[]{\label{FIG1d}
\scalebox{0.14}{\includegraphics{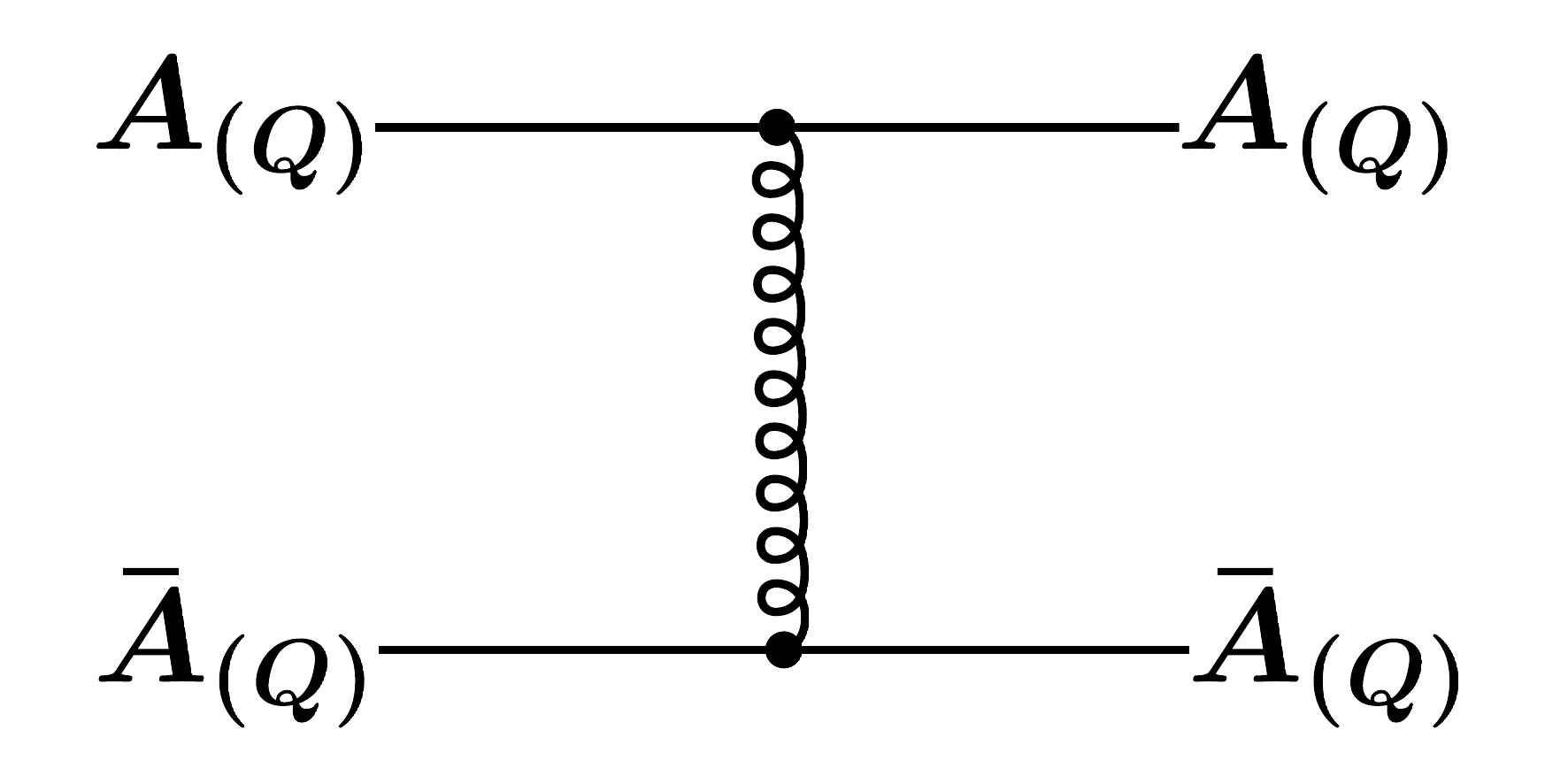}}}
\caption{The one-gluon-exchange interactions between diquark and antidiquark at the diquark level. $S_{(Q)}$ denotes the light (heavy) scalar diquark, and $A_{(Q)}$ the light (heavy) axial vector diquark. $S_{(Q)}$ stands for $S$ or $S_Q$, and $A_{(Q)}$ for $A$ or $A_Q$.}
\label{FIG1}
\end{figure}

With the preparation above, we can calculate the one-gluon-exchange potentials between a diquark and an antidiquark, which correspond to the Feynman diagrams in FIG.~\ref{FIG1}. Generally, the scattering amplitude $\mathrm{\mathcal{M}}$ is related to the interaction potential in the momentum space in terms of the Breit equation
\begin{eqnarray}
\mathcal{V}(q)=\frac{\mathcal{M}(q)}{\sqrt{\prod_{i}2m_{i}\prod_{f}2m_{f}}},\label{eq43}
\end{eqnarray}
where the $m_{i}$ and $m_{f}$ denote the masses of particle in initial and final states , respectively. The potential in the coordinate space $V(r)$ is obtained after performing the Fourier transformation
\begin{eqnarray}
V(r)=\int\frac{d^{3}q}{(2\pi)^{3}}e^{i\boldsymbol{\mathit{q}}\cdot \boldsymbol{\mathit{r}}}V(q).\label{eq44}
\end{eqnarray}

The color singlet wave function of diquark and antidiquark reads
\begin{eqnarray}
\chi_C&=&\frac{1}{\sqrt{3}}(\chi_{rg}\chi_{\bar{r}\bar{g}}+\chi_{rb}\chi_{\bar{r}\bar{b}}+\chi_{gb}\chi_{\bar{g}\bar{b}}).
\end{eqnarray}
And the expressions of the effective potentials are obtained as
\begin{eqnarray}
    V_{a}^{(Q)}(r)&=&-\frac{g^{2}}{3\pi}\frac{1}{r}-\frac{g^{2}}{3m_{S_{(Q)}}^{2}}\delta^{3}(r),\label{eq46} \\ 
    V_{b}^{(Q)}(r)&=&-\frac{g^{2}}{3\pi}\frac{1}{r}\boldsymbol\epsilon_{2}\cdot\boldsymbol\epsilon_{4}^{\dagger}-\frac{2g^{2}}{9m_{S_{(Q)}}m_{A_{(Q)}}}\boldsymbol\epsilon_{2}\cdot\boldsymbol\epsilon_{4}^{\dagger}\delta^{3}(r)\nonumber\\ 
    &&-\frac{g^{2}}{12\pi m_{S_{(Q)}}m_{A_{(Q)}}}\frac{1}{r^3}S(\hat{\boldsymbol{\mathit{r}}},\boldsymbol\epsilon_2,\boldsymbol\epsilon_4^\dag),\label{eq47}  \\ 
    V_{c}^{Q}(r)&=&-\frac{8}{9}d_{1Q}^{2}\boldsymbol\epsilon_{2}\cdot\boldsymbol\epsilon_{3}^{\dag}\delta^{3}(r)-\frac{d_{1Q}^{2}}{3\pi}\frac{1}{r^3}S(\hat{\boldsymbol{\mathit{r}}},\boldsymbol\epsilon_2,\boldsymbol\epsilon_3^\dag),\label{eq48} \\ 
    V_c(r)&=&0,\label{eq49}\\
    V_{d}^{(Q)}(r)&=&-\frac{g^{2}}{3\pi}\frac{\boldsymbol\epsilon_{1}\cdot\boldsymbol\epsilon_{3}^{\dagger}\boldsymbol\epsilon_{2}\cdot\boldsymbol\epsilon_{4}^{\dagger}}{r}+\frac{1}{9m_{A_{(Q)}}^{2}}\left[-g^{2}\boldsymbol\epsilon_{1}\cdot\boldsymbol\epsilon_{3}^{\dagger}\boldsymbol\epsilon_{2}\cdot\boldsymbol\epsilon_{4}^{\dagger}\right. \nonumber\\ 
    &&-(2d_{2(Q)}g+2d_{2(Q)}^{2}+g^{2})\boldsymbol\epsilon_{1}\cdot\boldsymbol\epsilon_{2}\boldsymbol\epsilon_{3}^{\dagger}\cdot\boldsymbol\epsilon_{4}^{\dagger}\nonumber\\ &&\left.+2(d_{2(Q)}^{2}+d_{2(Q)}g)\boldsymbol\epsilon_{1}\cdot\boldsymbol\epsilon_{4}^{\dagger}\boldsymbol\epsilon_{2}\cdot\boldsymbol\epsilon_{3}^{\dagger}\right]\delta^{3}(r)  \nonumber\\ 
    &&+\frac{1}{12\pi m_{A_{(Q)}}^{2}}\frac{1}{r^{3}}\left[-g^{2}\boldsymbol\epsilon_2\cdot\boldsymbol\epsilon_4^\dag S(\hat{\boldsymbol{\mathit{r}}},\boldsymbol\epsilon_1,\boldsymbol\epsilon_3^\dag)\right.\nonumber\\  
    &&-g^{2}\boldsymbol\epsilon_1\cdot\boldsymbol\epsilon_3^\dag S(\hat{\boldsymbol{\mathit{r}}},\boldsymbol\epsilon_2,\boldsymbol\epsilon_4^\dag)+d_{2(Q)}^{2}\boldsymbol\epsilon_{1}\cdot\boldsymbol\epsilon_{2}S(\hat{\boldsymbol{\mathit{r}}},\boldsymbol\epsilon_3^\dag,\boldsymbol\epsilon_4^\dag)\nonumber\\     
    &&+(d_{2(Q)}+g)^{2}\boldsymbol\epsilon_{3}^{\dagger}\cdot\boldsymbol\epsilon_{4}^{\dagger}S(\hat{\boldsymbol{\mathit{r}}},\boldsymbol\epsilon_1,\boldsymbol\epsilon_2)\nonumber\\    
    &&-(d_{2(Q)}^{2}+d_{2(Q)}g)\boldsymbol\epsilon_{1}\cdot\boldsymbol\epsilon_{4}^{\dagger}S(\hat{\boldsymbol{\mathit{r}}},\boldsymbol\epsilon_2,\boldsymbol\epsilon_3^\dag) \nonumber\\ 
    && \left.-(d_{2(Q)}^{2}+d_{2(Q)}g)\boldsymbol\epsilon_{2}\cdot\boldsymbol\epsilon_{3}^{\dagger}S(\hat{\boldsymbol{\mathit{r}}},\boldsymbol\epsilon_1,\boldsymbol\epsilon_4^\dag) \right]. \label{eq50}
\end{eqnarray}
In the above equations, the subscripts $a,b,c,d$ represents the labels in FIG.~\ref{FIG1}. The potential with (without) superscript $Q$ means it corresponds to heavy (light) diquarks' interaction. The superscript $(Q)$ means the potentials of heavy diquarks' interaction and light diquarks' interaction are written together. For example, $V^{(Q)}_a(r)$ means $V_a(r)$ or $V^Q_a(r)$. $m_{S}$, $m_{S_{Q}}$, $m_{A}$ and $m_{A_{Q}}$ are the masses of diquarks $S_{qq}$, $S_{Qq}$, $A_{qq}$ and $A_{Qq}$, respectively. $S(\hat{\boldsymbol{\mathit{r}}},\boldsymbol{A},\boldsymbol{B})=3(\boldsymbol{A}\cdot \hat{\boldsymbol{\mathit{r}}})(\boldsymbol{B}\cdot \hat{\boldsymbol{\mathit{r}}})-\boldsymbol{A}\cdot \boldsymbol{B}$ and $\hat{\boldsymbol{\mathit{r}}}=\frac{\boldsymbol{\mathit{r}}}{r}$. We see that there is no tenser term in the potential for scalar diquark and scalar antidiquark. The Coulomb-type term in Eqs.~\eqref{eq47} and $\eqref{eq50}$ are proportional to $\boldsymbol\epsilon_{2}\cdot\boldsymbol\epsilon_{4}^{\dagger}$ and $\boldsymbol\epsilon_{1}\cdot\boldsymbol\epsilon_{3}^{\dagger}\boldsymbol\epsilon_{2}\cdot\boldsymbol\epsilon_{4}^{\dagger}$, respectively. After projecting them to different spins using $\langle 1, m'| \boldsymbol\epsilon_2\cdot \boldsymbol\epsilon_4^\dag |1, m\rangle=\delta_{m',m}$ and $\langle j, m'| \boldsymbol\epsilon_1\cdot \boldsymbol\epsilon_3^\dag \boldsymbol\epsilon_2\cdot \boldsymbol\epsilon_4^\dag |j, m\rangle=\delta_{m',m}$ with $j=0,1,2$, we can see that the Coulomb-type term is spin-independent. Due to the spin wave functions of the initial and final states corresponding to FIG.~\ref{FIG1c} are orthogonal, Eq. \eqref{eq48} does not involve Coulomb-type term. And the non-zero potentials in Eqs. \eqref{eq46}-\eqref{eq50} contain the contact terms. Besides, the non-zero potentials in Eqs. \eqref{eq47}-\eqref{eq50} all have tensor terms which are proportional to $\frac{1}{r^{3}}$. 

Note that the calculation is performed under the on‑shell approximation, where all potentials are independent of the gauge parameter in the gluon propagator. Taking Fig.~\ref{FIG1a} as an example, the amplitude 
\begin{eqnarray}
    \mathcal{M}_a^{(Q)}&\propto&(p_1+p_3)_\mu \frac{g^{\mu\nu}-(1-\xi)q^\mu q^\nu/q^2}{q^2+i\epsilon}(p_2+p_4)_\nu\nonumber\\
    &=&\frac{(p_1+p_3)\cdot(p_2+p_4)}{q^2+i\epsilon}.
\end{eqnarray}
We see that the term proportional to $\xi$ vanishes, since $(p_1+p_3)_\mu q^\mu=(p_1+p_3)_\mu (p_1-p_3)^\mu=p_1^2-p_3^2=0$. For the off‑shell case, it is necessary to extend the gauge symmetry to BRST symmetry, under which the ghost field is introduced to abrogate the dependence on the gauge parameter. 

\section{Determination of the coupling constants}\label{IV}
In the Lagrangians~\eqref{eq31}-\eqref{eq37}, the coupling constants $g$, $d_{1Q}$, $d_2$ and $d_{2Q}$ are still unknown. In order to determine them, we calculate the effective potentials in FIG.~\ref{FIG1} using Godfrey-Isgur quark (GI) model \cite{Godfrey:1985xj}. Then we compare them with those obtained from the Lagrangians to determine the coupling constants.

The GI model gives the potential between two quarks. In the present work, since our goal is to determine the coupling constants $g$, $d_{1Q}$, $d_2$ and $d_{2Q}$ in the effective Lagrangians, we only focus on the Coulomb and tensor force parts, i.e.,
\begin{eqnarray}
    V_{ij}(p,r)&=&\frac{\alpha_{s}(r)}{r}\textbf{\textit{F}}_{i}\cdot\textbf{\textit{F}}_{j}-\frac{\alpha_{s}(r)}{m_{i}m_{j}}\left[\frac{8\pi}{3}\boldsymbol{\mathit{S}}_{i}\cdot \boldsymbol{\mathit{S}}_{j}\delta^{3}(r)\right. \nonumber \\ 
    &&\left.+\frac{1}{r^{3}}\left(\frac{3\boldsymbol{\mathit{S}}_{i}\cdot \boldsymbol{\mathit{r}}\boldsymbol{\mathit{S}}_{j}\cdot \boldsymbol{\mathit{r}}}{r^{2}}-\boldsymbol{\mathit{S}}_{i}\cdot \boldsymbol{\mathit{S}}_{j}\right)\right]\textbf{\textit{F}}_{i}\cdot\textbf{\textit{F}}_{j}.
\end{eqnarray}
Here, $\textbf{\textit{F}}_{i}$ is defined as
\begin{eqnarray}
    \textbf{\textit{F}}_{i}=\left\{ 
    \begin{array}{lc}
        \frac{\lambda{_{i}}}{2}, & \rm{for\ quarks} \\
        \frac{\lambda{_{i}^{c}}}{2}=-\frac{\lambda_{i}^{*}}{2}, & \rm{for\ antiquarks}\\
    \end{array}\right. 
\end{eqnarray}
and $\alpha_{s}(r)=\sum\limits_{k}\alpha_{k}\frac{2}{\sqrt{\pi}}\int_{0}^{\gamma_{k}r}e^{-x^{2}}dx$, with $\alpha_{1,2,3}=0.25,0.15,0.2$ and $\gamma_{1,2,3}=\frac{1}{2},\frac{\sqrt{10}}{2},\frac{\sqrt{1000}}{2}$.

At the quark level, the color structure of diquark-antidiquark is $\bar{3}\otimes3$, so the color singlet wave function reads
\begin{eqnarray}
\chi_C
&=&\frac{1}{\sqrt{12}}\epsilon_{ijk}u_{1j}u_{2k}\epsilon_{ij^\prime k^\prime}\bar{u}_{3j^\prime}\bar{u}_{4k^\prime}\nonumber\\
&=&\frac{1}{\sqrt{12}}(u_{1j}u_{2k}\bar{u}_{3j}\bar{u}_{4k}-u_{1j}u_{2k}\bar{u}_{3k}\bar{u}_{4j}),
\end{eqnarray}
where $u_{ai}=r, g, b$ for $i=1, 2, 3$ with $a=1,2,3,4$ the label of quark, and the Einstein summation convention is used. The total spin wave functions of diquark and antidiquark systems are 
\begin{eqnarray}
|S_{(Q)}S_{(Q)}\rangle_{0,0}&=&\chi_{0,0}\chi_{0,0},\\
|S_{(Q)}A_{(Q)}\rangle_{1,M}&=&\chi_{0,0}\chi_{1,M},\\
|A_{(Q)}A_{(Q)}\rangle_{S,M}&=&\sum_{m,m^\prime}\langle 1,m;1,m^\prime|S,M\rangle\chi_{1,m}\chi_{1,m^\prime},
\end{eqnarray}
where
\begin{eqnarray}
\chi_{0,0}&=&\frac{1}{\sqrt{2}}(\uparrow\downarrow-\downarrow\uparrow),\\
\chi_{1,1}&=&\uparrow\uparrow,\\
\chi_{1,0}&=&\frac{1}{\sqrt{2}}(\uparrow\downarrow+\downarrow\uparrow),\\
\chi_{1,-1}&=&\downarrow\downarrow.
\end{eqnarray}

By comparing the Coulomb terms obtained from the effective Lagrangians and GI model, we have
\begin{eqnarray}
    \frac{g^2}{4\pi}&=&\alpha_{s}(r).
\end{eqnarray}
Considering the size of tetraquarks, we can approximately take $\alpha_{s}(r)\approx 0.6$, and henceforth the coupling $g$ is taken as $g\approx 2.7$.
If taking into account the $S_{(Q)}\bar{A}_{(Q)}\to A_{(Q)}\bar{S}_{(Q)}$ process in FIG. \ref{FIG1c}, the following relations are obtained
\begin{eqnarray}
    d_{1Q}^{2}&\approx&\frac{g^2}{16}\frac{1}{m_{q}^{2}}.
\end{eqnarray}
It is worth mentioning that the sign of the potential corresponding to $S_Q\bar{A}_Q\to A_Q\bar{S}_Q$ can not be fixed in GI model, which depends on the choice of the order of quarks. So we take the same sign as the potential obtained from the Lagrangians. For the $A_{(Q)}\bar{A}_{(Q)}$ system, the comparison of tensor terms obtained from the effective Lagrangians and GI model gives 
\begin{eqnarray}
    d_{2}^2&\approx&g^2\frac{m_{A_{qq}}^2}{m_q^2},\\
    d_{2Q}^2&\approx&\frac{g^2}{4}\frac{m_{A_{Qq}}^2}{m_q^2}.
\end{eqnarray}
The quark mass is $m_q=220$ MeV \cite{Godfrey:1985xj} and the diquark masses are $m_{A_{qq}}=840$ MeV, $m_{A_{cq}}=2138$ MeV, $m_{A_{bq}}=5465$ MeV \cite{Ferretti:2019zyh}. If the signs of $d_{1Q}$, $d_2$ and $d_{2Q}$ are choosing as positive, we get $d_{1Q}\approx 3.1\ \mathrm{GeV^{-1}}$, $d_{2}\approx 10.5$ and $d_{2Q}\approx 13.3\ (34.1)$ for charmed (bottomed) diquark.

\section{Summary}\label{V}  
In the present work, we construct a color gauge invariant theory of diquark interactions. Firstly, we show the transformation rule of the diquark fields and the definition of the covariant derivatives under SU(3) symmetry, and then construct the color gauge invariant Lagrangians describing the vertices of $S-S-G$, $A-A-G$, $S_Q-S_Q-G$, $A_Q-A_Q-G$ and $S_Q-A_Q-G$. Since in the flavour space, the matrix of $S$ is antisymmetric while $A$ and $G$ are symmetric, we get the Lagrangian $\mathcal{L}_{SAG}=0$ when performing the trace. Utilizing these Lagrangians, we derive the one-gluon-exchange effective potentials for diquark-antidiquark interactions.

By comparing the effective potentials with those obtained from the GI model, we determine the couplings of diquark-diquark-gluon, i.e., $g\approx 2.7$, $d_{1Q}\approx 3.1\ \mathrm{GeV^{-1}}$, $d_2\approx 10.5$ and $d_{2Q}\approx 13.3\ (34.1)$ for charmed (bottomed) diquark. We find that the $S_{(Q)}\bar{S}_{(Q)}$ potentials do not contain tensor term. For the $S_{(Q)}\bar{A}_{(Q)}$ potential, the tensor term is negligible comparing with the Coulomb term, due to the factor $\frac{1}{m_{S_{(Q)}}m_{A_{(Q)}}}$. For the process of $S\bar{A}\to A\bar{S}$, the corresponding potential is zero, while for $S_{Q}\bar{A}_{Q}\to A_{Q}\bar{S}_{Q}$, the potential involves contact and tensor terms. The $A_{(Q)}\bar{A}_{(Q)}$ potential is made of Coulomb, contact and tensor terms. We believe that our work will shed light on further studies of tetraquarks.

\section*{Acknowledgments}
We would like to thank Zi-Long Man for valuable discussion. This project is supported by the National Natural Science Foundation of China (NSFC) under Grants No. 11965016, 11705069 and 12335001, and the National Natural Science Foundation of China (Grant No. 12247101), the Fundamental Research Funds for the Central Universities (Grant No. lzujbky-2025-jdzx07), the Natural Science Foundation of Gansu Province (No. 22JR5RA389, No.25JRRA799), and the ‘111 Center’ under Grant No. B20063.

\section*{Appendix A: Equivalence of two different forms of $S_{(Q)}-A_{(Q)}-G$ terms}
Firstly, we prove the equivalence of $\epsilon^{\mu\nu\alpha\beta}\langle S_{Q}G_{\mu\nu}^{T}F_{Q\alpha\beta}^{\dagger}\rangle_{C}$ to $\epsilon^{\mu\nu\alpha\beta}\langle D_\mu S_{Q}G_{\nu\alpha}^{T}A_{Q\beta}^\dag\rangle_{C}$,
\begin{eqnarray}
&&\epsilon^{\mu\nu\alpha\beta}\langle S_{Q}G_{\mu\nu}^{T}F_{Q\alpha\beta}^{\dagger}\rangle_{C}\nonumber\\
&=&\epsilon^{\mu\nu\alpha\beta}\langle S_{Q}G_{\mu\nu}^{T}(D_\alpha A_{Q\beta}^{\dagger}-D_\beta A_{Q\alpha}^{\dagger})\rangle_{C}\nonumber\\
&=&2\epsilon^{\mu\nu\alpha\beta}\langle S_{Q}G_{\mu\nu}^{T}D_\alpha A_{Q\beta}^{\dagger}\rangle_{C}\nonumber
\end{eqnarray}
\begin{eqnarray}
&=&2\epsilon^{\mu\nu\alpha\beta}\langle S_{Q}G_{\mu\nu}^{T}(\partial_{\alpha}A_{Q\beta}^\dag+iA_{Q\beta}^\dag V_{\alpha}+iV_{\alpha}^{T}A_{Q\beta}^\dag+igA_{Q\beta}^\dag G_{\alpha}\nonumber\\
&&+igG_{\alpha}^{T}A_{Q\beta}^\dag)\rangle_{C}\nonumber \\
&=&2\epsilon^{\mu\nu\alpha\beta}\langle S_{Q}G_{\mu\nu}^{T}\partial_{\alpha}A_{Q\beta}^\dag+S_{Q}G_{\mu\nu}^{T}iA_{Q\beta}^\dag V_{\alpha}+S_{Q}G_{\mu\nu}^{T}iV_{\alpha}^{T}A_{Q\beta}^\dag\nonumber\\
&&+S_{Q}G_{\mu\nu}^{T}igA_{Q\beta}^\dag G_{\alpha}+S_{Q}G_{\mu\nu}^{T}igG_{\alpha}^{T}A_{Q\beta}^\dag\rangle_{C}\nonumber\\
&=&2\epsilon^{\mu\nu\alpha\beta}\langle S_{Q}G_{\mu\nu}^{T}\partial_{\alpha}A_{Q\beta}^\dag+iV_{\alpha}S_{Q}G_{\mu\nu}^{T}A_{Q\beta}^\dag +iS_{Q}V_{\alpha}^{T}G_{\mu\nu}^{T}A_{Q\beta}^\dag\nonumber\\
&&+igG_{\alpha}S_{Q}G_{\mu\nu}^{T}A_{Q\beta}^\dag +S_{Q}G_{\mu\nu}^{T}igG_{\alpha}^{T}A_{Q\beta}^\dag\rangle_{C}\nonumber\\
&\dot{=}&2\epsilon^{\mu\nu\alpha\beta}\langle -\partial_{\alpha}S_{Q}G_{\mu\nu}^{T}A_{Q\beta}^\dag-S_{Q}\partial_{\alpha}G_{\mu\nu}^{T}A_{Q\beta}^\dag+iV_{\alpha}S_{Q}G_{\mu\nu}^{T}A_{Q\beta}^\dag \nonumber\\
&&+iS_{Q}V_{\alpha}^{T}G_{\mu\nu}^{T}A_{Q\beta}^\dag+igG_{\alpha}S_{Q}G_{\mu\nu}^{T}A_{Q\beta}^\dag +S_{Q}G_{\mu\nu}^{T}igG_{\alpha}^{T}A_{Q\beta}^\dag\rangle_{C}\nonumber\\
&=&2\epsilon^{\mu\nu\alpha\beta}\langle -\partial_{\alpha}S_{Q}G_{\mu\nu}^{T}A_{Q\beta}^\dag+iV_{\alpha}S_{Q}G_{\mu\nu}^{T}A_{Q\beta}^\dag+iS_{Q}V_{\alpha}^{T}G_{\mu\nu}^{T}A_{Q\beta}^\dag \nonumber\\
&&+igG_{\alpha}S_{Q}G_{\mu\nu}^{T}A_{Q\beta}^\dag -S_{Q}[\bar{D}_{\alpha},G_{\mu\nu}^{T}]A_{Q\beta}^\dag+S_{Q}igG_{\alpha}^{T}G_{\mu\nu}^{T}A_{Q\beta}^\dag\rangle_{C}\nonumber\\
&=&2\epsilon^{\mu\nu\alpha\beta}\langle -\partial_{\alpha}S_{Q}G_{\mu\nu}^{T}A_{Q\beta}^\dag+iV_{\alpha}S_{Q}G_{\mu\nu}^{T}A_{Q\beta}^\dag+iS_{Q}V_{\alpha}^{T}G_{\mu\nu}^{T}A_{Q\beta}^\dag \nonumber\\
&&+igG_{\alpha}S_{Q}G_{\mu\nu}^{T}A_{Q\beta}^\dag +\frac{i}{g}S_{Q}[\bar{D}_{\alpha},[\bar{D}_{\mu},\bar{D}_{\nu}]]A_{Q\beta}^\dag\nonumber\\
&&+S_{Q}igG_{\alpha}^{T}G_{\mu\nu}^{T}A_{Q\beta}^\dag\rangle_{C}\nonumber\\
&=&2\epsilon^{\mu\nu\alpha\beta}\langle -\partial_{\alpha}S_{Q}G_{\mu\nu}^{T}A_{Q\beta}^\dag+iV_{\alpha}S_{Q}G_{\mu\nu}^{T}A_{Q\beta}^\dag+iS_{Q}V_{\alpha}^{T}G_{\mu\nu}^{T}A_{Q\beta}^\dag \nonumber\\
&&+igG_{\alpha}S_{Q}G_{\mu\nu}^{T}A_{Q\beta}^\dag+S_{Q}igG_{\alpha}^{T}G_{\mu\nu}^{T}A_{Q\beta}^\dag\rangle_{C}\nonumber\\
&=&-2\epsilon^{\mu\nu\alpha\beta}\langle(\partial_{\alpha}S_{Q}-iV_{\alpha}S_{Q}-iS_{Q}V_{\alpha}^{T} -igG_{\alpha}S_{Q}\nonumber\\
&&-igS_{Q}G_{\alpha}^{T})G_{\mu\nu}^{T}A_{Q\beta}^\dag\rangle_{C}\nonumber\\
&=&-2\epsilon^{\mu\nu\alpha\beta}\langle D_\alpha S_{Q}G_{\mu\nu}^{T}A_{Q\beta}^\dag\rangle_{C}\nonumber\\
&=&-2\epsilon^{\mu\nu\alpha\beta}\langle D_\mu S_{Q}G_{\nu\alpha}^{T}A_{Q\beta}^\dag\rangle_{C},
\end{eqnarray}
where $\bar{D}_\alpha=\partial_\alpha+igG_\alpha^T$, the symbol $\dot{=}$ means the surface term is neglected, and the relations $\epsilon^{\mu\nu\alpha\beta}[\bar{D}_{\alpha},[\bar{D}_{\mu},\bar{D}_{\nu}]]=0$ and $[\bar{D}_\mu,\bar{D}_\nu]=igG_{\mu\nu}^T$ are used.

\section*{Appendix B: Proof of $\mathcal{L}_{SAG}=0$}
Next, we give the proof that the vertex of $S-A-G$ is zero,
\begin{eqnarray}
\mathcal{L}_{SAG}&=&d_1\epsilon^{\mu\nu\alpha\beta}\langle D_{\mu}SG_{\nu\alpha}^{T}A_{\beta}^{\dagger}+A_{\beta}G_{\nu\alpha}^{T}D_{\mu}S^{\dagger}\rangle_{CF}\nonumber\\
&=&d_1\epsilon^{\mu\nu\alpha\beta}\langle (D_{\mu}S)^{ij}(G_{\nu\alpha}^{T})^{jk}A_{\beta}^{\dagger ki}\rangle_C+h.c.\nonumber\\
&=&d_1\epsilon^{\mu\nu\alpha\beta}\langle -(D_{\mu}S)^{ji}(G_{\nu\alpha}^{T})^{kj}A_{\beta}^{\dagger ik}\rangle_C+h.c.\nonumber\\
&=&d_1\epsilon^{\mu\nu\alpha\beta}\langle -(D_{\mu}S)^{ji})(A_{\beta}^{\dagger}G_{\nu\alpha}^{T})^{ij}\rangle_C+h.c.\nonumber\\
&=&d_1\epsilon^{\mu\nu\alpha\beta}\langle -(D_{\mu}S)^{ji})(G_{\nu\alpha}^{T}A_{\beta}^{\dagger})^{ij}\rangle_C+h.c.\nonumber\\
&=&d_1\epsilon^{\mu\nu\alpha\beta}\langle -(D_{\mu}S)^{ji}(G_{\nu\alpha}^{T})^{ik}A_{\beta}^{\dagger kj}\rangle_C+h.c.\nonumber\\
&=&d_1\epsilon^{\mu\nu\alpha\beta}\langle -D_{\mu}SG_{\nu\alpha}^{T}A_{\beta}^{\dagger}\rangle_{CF}+h.c.\nonumber\\
&=&-\mathcal{L}_{SAG}\nonumber\\
&=&0,\nonumber
\end{eqnarray}
where the superscripts $i,j,k$ denote the flavour indices, and the relation $(D_{\mu}S)^{ij}=-(D_{\mu}S)^{ji}$ is used in the third step.

\section*{Appendix C: The form factor of diquark-diquark-gluon vertices}
In the derivation of Eqs. \eqref{eq46}-\eqref{eq50}, we do not consider the internal structure of diquarks, i.e., the diquarks are treated as point-like particles. However, the impact of spatial extension of diquarks needs to be considered. Similar to Ref.~\cite{Anselmino:1987vk,Kroll:1987pj,Kroll:1988cd}, we take the form factor $F(q^2)$ to describe the internal structure of diquarks, 
\begin{eqnarray}
    F(q^2)&=&\frac{\alpha_s(q^2)Q_0^2}{Q_0^2-q^2}.
\end{eqnarray}
Here, we take $\alpha_s(q^2)$ as a parameter, $\alpha_s(q^2)\approx0.6$ and $Q_0^2=3.2$ GeV$^2$.

After consider the form factor $F(q^2)$, the effective potentials in coordinate space of FIG.~1(a)-(d) are obtained as follows
\begin{eqnarray}
    V_{a}^{(Q)}(r)&=&-\frac{4}{3}g^{2}\alpha_s^2Y(Q_0,r)+\frac{g^{2}\alpha_s^2}{3m_{S_{(Q)}}^{2}}Z(Q_0,r), \\ 
    V_{b}^{(Q)}(r)&=&-\frac{4}{3}g^{2}\alpha_s^2\boldsymbol\epsilon_{2}\cdot\boldsymbol\epsilon_{4}^{\dagger}Y(Q_0,r)+\frac{2}{9m_{S_{(Q)}}m_{A_{(Q)}}}g^{2}\alpha_s^2\boldsymbol\epsilon_{2}\cdot\boldsymbol\epsilon_{4}^{\dagger}\nonumber\\ 
    &&\times Z(Q_0,r)-\frac{2}{9 m_{S_{(Q)}}m_{A_{(Q)}}}g^{2}\alpha_s^2S(\hat{\boldsymbol{\mathit{r}}},\boldsymbol\epsilon_2,\boldsymbol\epsilon_4^\dag)T(Q_0,r),  \nonumber\\\\ 
    V_{c}^{Q}(r)&=&\frac{8}{9}d_{1Q}^{2}\alpha_s^2\boldsymbol\epsilon_{2}\cdot\boldsymbol\epsilon_{3}^{\dag}Z(Q_0,r)-\frac{4}{9}d_{1Q}^{2}\alpha_s^2S(\hat{\boldsymbol{\mathit{r}}},\boldsymbol\epsilon_2,\boldsymbol\epsilon_3^\dag)\nonumber\\
    &&\times T(Q_0,r), \\ 
    V_c(r)&=&0,\\
    V_{d}^{(Q)}(r)&=&-\frac{4}{3}g^{2}\alpha_s^2\boldsymbol\epsilon_{1}\cdot\boldsymbol\epsilon_{3}^{\dagger}\boldsymbol\epsilon_{2}\cdot\boldsymbol\epsilon_{4}^{\dagger}Y(Q_0,r)+\frac{\alpha_s^2}{9m_{A_{(Q)}}^{2}}\left[g^{2}\boldsymbol\epsilon_{1}\cdot\boldsymbol\epsilon_{3}^{\dagger}\right. \nonumber\\ 
    &&\times\boldsymbol\epsilon_{2}\cdot\boldsymbol\epsilon_{4}^{\dagger}+(2d_{2(Q)}g+2d_{2(Q)}^{2}+g^{2})\boldsymbol\epsilon_{1}\cdot\boldsymbol\epsilon_{2}\boldsymbol\epsilon_{3}^{\dagger}\cdot\boldsymbol\epsilon_{4}^{\dagger}\nonumber\\ &&\left.-2(d_{2(Q)}^{2}+d_{2(Q)}g)\boldsymbol\epsilon_{1}\cdot\boldsymbol\epsilon_{4}^{\dagger}\boldsymbol\epsilon_{2}\cdot\boldsymbol\epsilon_{3}^{\dagger}\right]Z(Q_0,r)  \nonumber\\ 
    &&-\frac{1}{9 m_{A_{(Q)}}^{2}}\alpha_s^2\left[g^{2}\boldsymbol\epsilon_2\cdot\boldsymbol\epsilon_4^\dag S(\hat{\boldsymbol{\mathit{r}}},\boldsymbol\epsilon_1,\boldsymbol\epsilon_3^\dag)\right.\nonumber\\  
    &&+g^{2}\boldsymbol\epsilon_1\cdot\boldsymbol\epsilon_3^\dag S(\hat{\boldsymbol{\mathit{r}}},\boldsymbol\epsilon_2,\boldsymbol\epsilon_4^\dag)-d_{2(Q)}^{2}\boldsymbol\epsilon_{1}\cdot\boldsymbol\epsilon_{2}S(\hat{\boldsymbol{\mathit{r}}},\boldsymbol\epsilon_3^\dag,\boldsymbol\epsilon_4^\dag)\nonumber\\     
    &&-(d_{2(Q)}+g)^{2}\boldsymbol\epsilon_{3}^{\dagger}\cdot\boldsymbol\epsilon_{4}^{\dagger}S(\hat{\boldsymbol{\mathit{r}}},\boldsymbol\epsilon_1,\boldsymbol\epsilon_2)\nonumber\\    
    &&+(d_{2(Q)}^{2}+d_{2(Q)}g)\boldsymbol\epsilon_{1}\cdot\boldsymbol\epsilon_{4}^{\dagger}S(\hat{\boldsymbol{\mathit{r}}},\boldsymbol\epsilon_2,\boldsymbol\epsilon_3^\dag) \nonumber\\ 
    && \left.+(d_{2(Q)}^{2}+d_{2(Q)}g)\boldsymbol\epsilon_{2}\cdot\boldsymbol\epsilon_{3}^{\dagger}S(\hat{\boldsymbol{\mathit{r}}},\boldsymbol\epsilon_1,\boldsymbol\epsilon_4^\dag) \right]T(Q_0,r).
\end{eqnarray}
In the above effective potentials, 
\begin{eqnarray}
Y(Q_0,r)&=&\frac{1}{4\pi r}(1-e^{-Q_0 r})-\frac{Q_0}{8\pi}e^{-Q_0 r},\\
Z(Q_0,r)&=&\nabla^2Y(Q_0,r),\\
T(Q_0,r)&=&r\frac{\partial}{\partial r}\frac{1}{r}\frac{\partial}{\partial r}Y(Q_0,r).
\end{eqnarray}

\bibliographystyle{apsrev4-1}
\bibliography{ref}

\end{document}